\documentclass[helvet,twocolumn]{NaudLab-Class}
\usepackage{blindtext}
 
\usepackage{mdframed}
\definecolor{shadecolor}{RGB}{220,220,220}

\usepackage{xcolor}

\def\mathcolor#1#{\@mathcolor{#1}}
\def\@mathcolor#1#2#3{%
  \protect\leavevmode
  \begingroup
    \color#1{#2}#3%
  \endgroup
}

\usepackage{abstract}

\mdfdefinestyle{mdfabstract}{%
    linecolor=blue, linewidth=0.8pt,
    backgroundcolor=shadecolor,
    leftline=false, rightline=false,
    topline=false,
    bottomline=false,
    innertopmargin=0.25cm, innerbottommargin=0.25cm,
    innerleftmargin=0.25cm, innerrightmargin=0.25cm,
}
\leadauthor{Naud}

\begin{document}

\title{Silences, Spikes and Bursts: Three-Part Knot of the Neural Code}
\shorttitle{Bursts – Three-Part Knot of the Neural Code}

\author[1,2,3,4,\Letter]{Richard Naud}
\affil[1]{Department of Cellular and Molecular Medicine, University of Ottawa, K1H 8M5, Ottawa, Canada}
\affil[2]{Department of Physics, University of Ottawa, K1H 8M5, Ottawa, Canada}
\affil[3]{Center for Neural Dynamics, University of Ottawa, K1H 8M5, Ottawa, Canada}
\affil[4]{Brain and Mind Institute, University of Ottawa, K1H 8M5, Ottawa, Canada}

\author[2,3,4]{Zachary Friedenberger}
\author[1,3,4]{Katalin Toth}

\maketitle
\begin{mdframed}[style=mdfabstract]
        \begin{onecolabstract}
            \noindent 
When a neuron breaks silence, it can emit action potentials in a number of patterns. Some responses are so sudden and intense that electrophysiologists felt the need to single them out, labeling action potentials emitted at a particularly high frequency with a metonym -- bursts. Is there more to bursts than a figure of speech? After all, sudden bouts of high-frequency firing are expected to occur whenever inputs surge. The burst coding hypothesis advances that the neural code has three syllables: silences, spikes and bursts. We review evidence supporting this ternary code in terms of devoted mechanisms for burst generation, synaptic transmission and synaptic plasticity. We also review the  learning and attention theories for which such a triad is beneficial.
        \end{onecolabstract}
\end{mdframed}


\begin{corrauthor}
\texttt{rnaud\at uottawa.ca}
\end{corrauthor}

\section{Introduction}

\begin{figure*}[h!]
\centering
\includegraphics[width=1\textwidth]{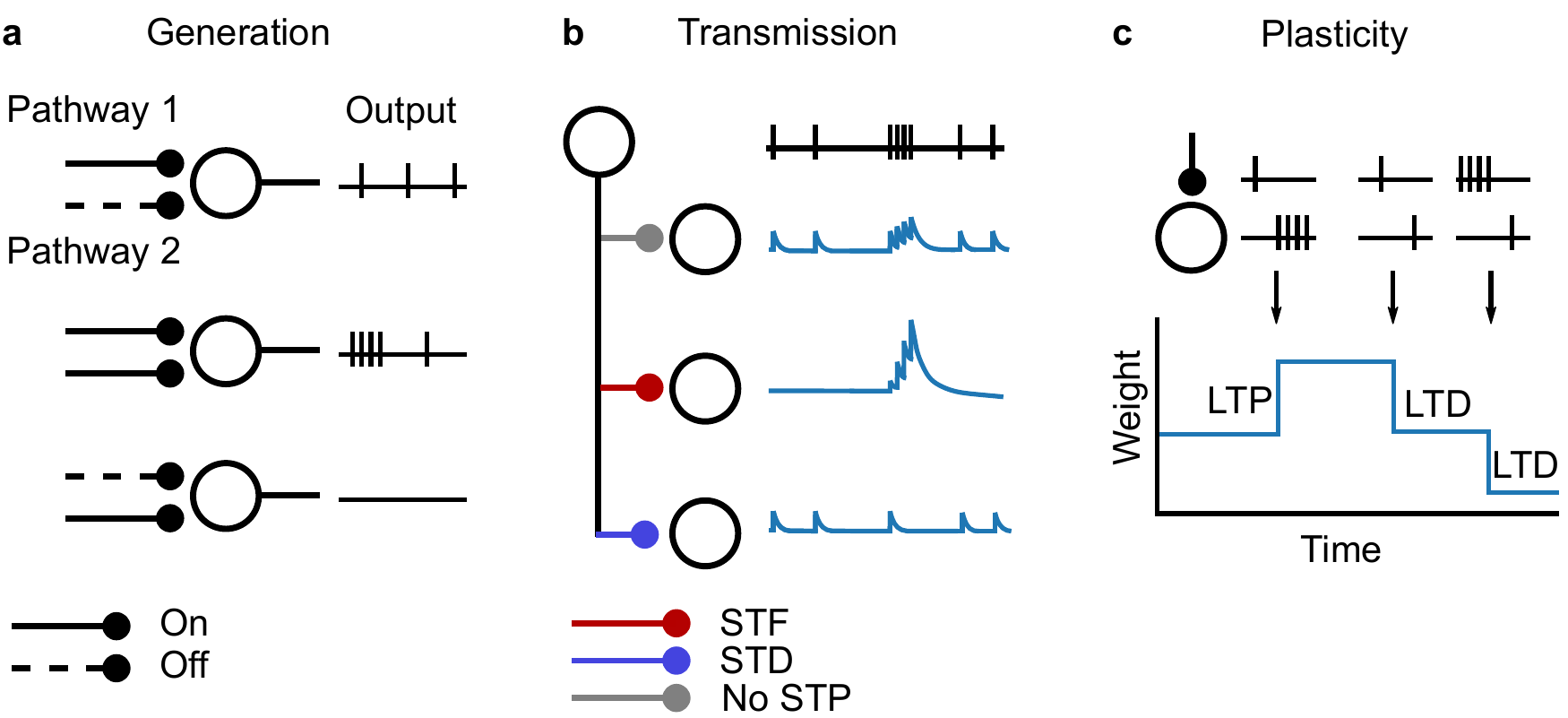}
\caption{Burst coding hypothesis(es). \textbf{a} Burst coding arising from the observation of response patterns arising only when a specific generation mechanism is present. Here a synaptic pathway (pathway 1) engages only spikes in relative isolation, and another synaptic pathway (pathway 2) engages bursts when present in conjunction with pathway 1. \textbf{b} Burst coding also arises from the observation that STD and STF synapses transmit bursts differently, with STD transmitting isolated spikes and the first spikes in a burst, and STF transmitting mostly the later spikes in a burst. \textbf{c} Burst coding surfaces from long-term potentiation, where pairing pre-post activity with bursts leads to LTP while pairing with isolated spikes leads to LTD.}
\label{fig:variations}
\end{figure*}
The problem of neural coding is to attribute the correct interpretation to neuronal signals. The "\emph{basis of sensation}" is generally that neurons represent input features with the number but not the shape of action potentials \cite{Adrian1928a}. Its central dogma stated that "\emph{high impulse frequency} (...) \emph{corresponds to high certainty that the trigger feature is present}" \cite{Barlow1972a}. In the presence of noise, interpreting neuronal responses becomes a statistical problem whereby high impulse frequencies may be randomly generated and only imply the presence of a feature if these high frequencies occur consistently. It has been argued that, somewhat counter-intuitively, such random utterance of all firing frequencies is a  way of maximizing information transmission \cite{Atick1990a}. As a result, inputs are represented by computing averages, turning the problem of neural coding towards the challenge of interpreting rate-based responses in large, possibly heterogeneous and context-dependent populations \cite{Georgopoulos1986a,Averbeck2006a,Shamir2006a,Yu2008a}. 

This dogma has been challenged by the idea that not all spikes are equal, that the central currency for neurons is not simply the spike, that the result is not a binary code of spikes and silences, but instead that the code is ternary: formed by either silences, spikes or bursts of high-frequency action potentials. In this burst coding hypothesis, high impulse frequency is not the sum of its parts, but a distinct type of event immersed in a separate causal chain. 

We review evidence supporting the burst coding hypothesis according to three of its variations (Fig. \ref{fig:variations}). The first variation asks if neurons have evolved special means to generate bursts, which are distinct from those to produce spikes in relative isolation. The second and third variations focus on the question of whether synapses treat bursts and spikes in the same way, either for transmission or for engaging in long-term plasticity. Each of these three hypotheses could be unrelated, but we show that a number of algorithms would benefit from an alignment between the generation mechanisms and the synaptic mechanisms. When all these hypotheses converge, the nervous system acquires the capability to process two streams of information simultaneously. Theories exploiting this capability for implementing top-down attention and learning have only begun to be considered.





\section{What is a burst?}

If the brain uses a ternary code, and when recording from neurons in vivo, should we expect to see well separated events? some events unfolding at 100 Hz and some other unfolding at much lower frequencies? While many neurons display such bimodal interspike interval (ISI) distributions, many do not. Does this mean that neurons displaying unimodal interspike interval distribution do not engage in burst coding? This question was addressed using information theory by Williams et al. (2021) \cite{Williams2021a}. The study found that, surprisingly, a unimodal distribution of intervals is not necessarily associated with a drop of information transmitted using a burst code. This is consistent with studies in the hippocampus that found well-defined complex spikes -- the very stereotype of a burst -- have ISIs ranging from 10 to 30 ms \cite{Epsztein2011a}, that is, frequencies produced by normal spiking during a place field. Consistently, as high-frequency events combine plateau potentials and normal firing, estimating burstiness based on plateau potentials leads to lower fractions than based on interspike intervals \cite{Bittner2015a,Sanders2019a}.   A unimodal interspike interval distribution implies that there is no logical interspike interval threshold that perfectly separate isolated spikes from bursts. Bursts can still be differentiated by including in addition to the interspike interval the number of intra-burst spikes and the presence of a sustained depolarization.

Invariably, the question must come: what exactly is a burst? If it is a sudden bout of firing at a high frequency, then precisely how high is that frequency and how sudden must the frequency increase be? How many spikes must be produced at high frequency? When does it end? Is an alteration of spike shape or the presence of a sustained depolarization relevant? Does the answer to any of these questions depend on cell type, area, or age? 
There is currently no consensus to answer these questions, but this is not to say that no attempts were made. Various metrics have been proposed. Some require a spike-per-spike classification of firing patterns \cite{Harris2001a,Naud2018a}, others do not \cite{Compte2003a,Shinomoto2009a,Simonnet2019a,Insanally2019a}. Yet other metrics measure the association between spike timing codes and external correlates without an estimate of burstiness nor a precise definition for bursting \cite{Kayser2009a,deKock2021a}. Some of these methods have parameters that must be set, while the others come with methodologies for self-tuning according to a particular objective (parameter free). Other than the information-theory metrics who evade the question, these metrics are implementations of a particular definition of bursting and assume somewhat different answers to the question “what is a burst?”. 

\subsection{Different together: when a burst is more than a bunch of spikes}

The burst coding hypothesis offers a loose definition: bursts are a family a firing patterns that trigger physiological mechanisms not engaged by the same number of spikes in relative isolation. Figure \ref{fig:variations} illustrates three variations of this question: generation, transmission and plasticity.

\begin{figure*}
    \centering
    \includegraphics{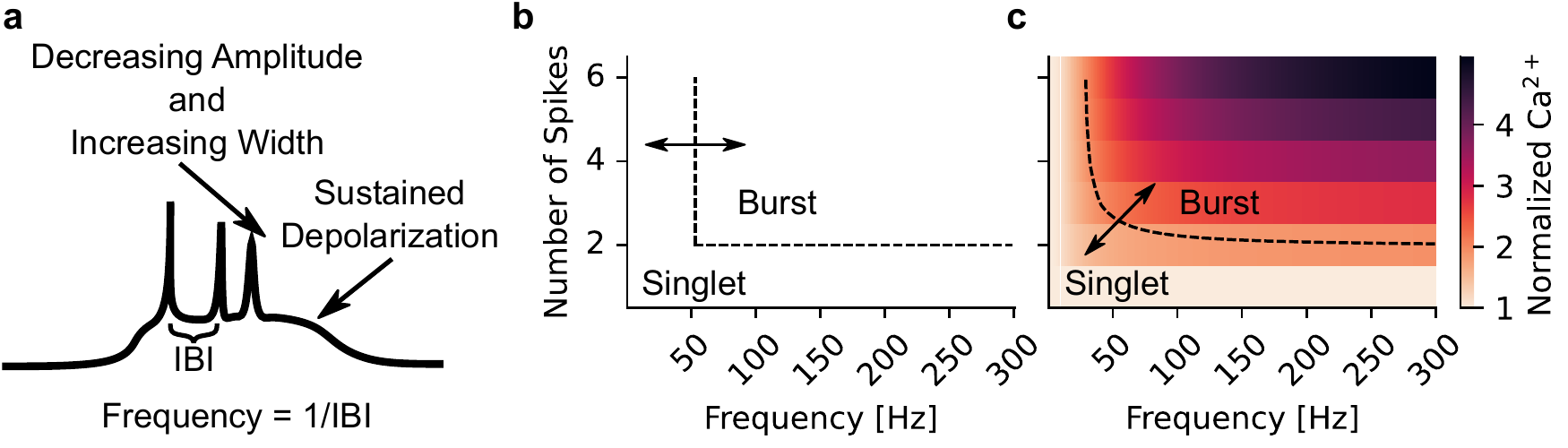}
    \caption{What is a burst? \textbf{a} Features associated with with bursting are high intra-burst frequencies, sustained depolarization, decreasing action potential amplitude and widening of action potential shape. \textbf{b} Defining bursting based on the number of spikes at a high frequency, here 2 spikes at 50 Hz is illustrated, but other choices of thresholds are possible. \textbf{c} Simulating calcium traces left by action potentials (exponential decay, 50 ms) and plotting the heatmap of peak calcium as generated by a number of spikes (y-axis) with a given frequency (x-axis). A threshold in peak calcium would correspond to a nonlineary curve in the space of frequency vs number of spikes, indicating a tradeoff between these two variable. }
    \label{fig:What}
\end{figure*}

Do neurons reserve some firing patterns for the response to specific types of stimulation? In all cells, short and long after-hyperpolarization currents following every action potential \cite{Schwindt1988a,Lundstrom2010a,Pozzorini2013a}. These combine with a moving threshold \cite{Azouz2000a,Mensi2012a,Harkin2023a} to prevent bouts of high-frequency action potentials to occur. Modelling studies have shown that that these properties ensure neurons responds rarely with high-frequency action potentials, although very large input fluctuations can transiently overcome these constraints \cite{Vogels2005a,Ostojic2011a}. Such random fluctuations are themselves constrained by feedforward and feedback inhibition, which ensures that any surge in excitation is followed by a calculated surge in inhibition, limiting depolarization in duration \cite{Pouille2001a,Swadlow2001a}. A number of cellular properties can evade these restrictions and produce high-frequency firing. These properties almost exclusively involve voltage gated calcium channels (VGCCs; for non-VGCC based bursting, see \cite{Haj1997a,Brumberg2000a,Doiron2007a}). In the thalamus, potent T-type VGCCs can de-inactivate following at least 100 ms of hyperpolarization and produce bursting upon release from inhibition \cite{Deschenes1984a,Huguenard1992a,Wang1991a}. In vitro studies have shown that these bursts are fast, with intra-burst intervals of 3-4 ms and last for less than 100 ms \cite{Deschenes1984a}. Duration of thalamic burst is variable and depends on the intensity of burst-generating inputs \cite{Mease2017a}. Dendritic VGCC in Purkinje cells lead to complex spikes \cite{Konnerth1992a,Davie2008a} having intraburst intervals of 2-5~ms with a sustained depolarization, alteration of spike shape and durations varying between 6 and 15 ms \cite{Maruta2007a,Yang2014a}, a duration that is controlled by inputs responsible for complex spike generation. In the hippocampus and cortex, VGCCs \cite{Williams2018a} in the dendrites of pyramidal cells of the cortex and hippocampus also generate dendrite-dependent VGCCs-based plateau-like potentials. In vitro studies indicate that these events contain spikes at a frequency close to 100 Hz, show alteration of spike shape, a sustained depolarization and last between 40 ms to 100 ms \cite{Kandel1961a,Larkum1999a,Larkum1999b,Takahashi2009a}. As for the cerebellum, duration and frequency is variable, with longer bouts being generated with higher input intensity and duration. Correspondingly, the presence of acetylcholine (ACh) give rise to longer (>500 ms) bouts of high-frequency firing \cite{Williams2018a}. In all areas, the bouts of high-frequency firing are associated with a gradual decrease of the amplitude of action potentials within the burst and a sustained depolarization. Together, neurons restrict the generation of bouts of high-frequency firing (above 100 Hz), events that are characterized by sustained depolarization and an alteration of spike shape with durations of at least 10~ms.

What type of stimulation patterns are transmitted differently by synapses? At some synapses, sudden bouts of high-frequency firing will either strongly and transiently depress or potentiate the post-synaptic effect of action potentials. This short-term plasticity (STP) can be frequency-dependent, and the effects can accumulate during a high-frequency train. There is a great diversity of STP expression that appears to depend on both pre- and post-synaptic cell classes. Some show a strong frequency dependence that gets engaged at a slow frequency ($1-10$ Hz) but is maximally expressed at high frequencies ($50-200$ Hz), others only start to be expressed at high frequencies, others show STP of opposite polarities at low and high-frequencies (\textit{i.e.} short-term facilitation (STF) and short-term depression (STD)), and yet others show no frequency-dependence at all \cite{Toth2000a,Dittman2000a,Salin1996a,Jackman2016a,Xu2012b,Lefort2009a,Campagnola2022a}.  A group of 6-8 high-frequency action potentials can induce up to 6-fold change in amplitude \cite{Toth2000a,Sun2009b,Xu2012b}. Such large change of post-synaptic amplitude can happen  for spikes in relative isolation (e.g. at mossy fiber synapses \cite{Toth2000a}), but facilitation of isolated firing is uncommon at most synapses. Adding a sustained depolarization to the high-frequency event does not alter transmission further \cite{Apostolides2016a}.  Thus the properties of STP do not provide a precise definition, but indicate that we may consider firing above 100 Hz a burst, with the possibility of trading lower firing frequency with greater number of spikes.

To complete the picture, we may look into the firing patterns that trigger a form of long-term synaptic plasticity not engaged with spikes in relative isolation. Alongside neuromodulation \cite{He2007a} and pre-post spike timing \cite{Markram1997a}, firing patterns are a central factor determining the relative strength of long-term potentiation (LTP) with respect to long-term depression (LTD) \cite{Sjostrom2001a,Pfister2006a,Lisman2005a,Froemke2006a}. In the cerebellum, LTD of parallel fiber synapses is induced by pairing parallel fibers inputs with complex-spike-inducing inputs \cite{Ito1982a,Suvrathan2016a}, parallel fiber inputs alone induce LTP, an observation that is thought to generalize to pairing parallel fiber stimulation with weak levels of activity of the Purkinje cell  \cite{Lev2002a,Coesmans2004a,Jorntell2006a}.  In the hippocampus and cortex, it is LTP that is induced by pairing pre-synaptic activity with high-frequency bursting post-synaptic, while isolated post-synaptic spikes express LTD \cite{Sjostrom2001a,Lisman2005a,Inglebert2020a,Froemke2006a,Bittner2017a}. In vitro studies have indicatd that LTP is expressed by frequencies of at least 40-50 Hz \cite{Sjostrom2001a,Pfister2006a}, but these studies with higher concentration of extracellular calcium than is to be expected in the living brain. In more physiological conditions, frequencies of at least 60 Hz are required and longer bursts are more likely to express LTP than shorter bursts \cite{Inglebert2020a}. For bursts longer than 2-3 spikes, burst duration reliably engages LTP. Further changes in burst duration as well as the presence of sustained depolarization can affect the learning rate and the duration of the eligibility trace \cite{Bittner2017a}. In these studies, however, the relative timing as well as the presence of neuromodulation can alter such first-order association between firing patterns and plasticity.

Looking for the firing pattern associated with specific mechanisms, we find, that events of more than 3 spikes at frequencies higher than 100 Hz can be labeled a burst in the hippocampus and cortex, and that a higher frequency may be required in the thalamus and the cerebellum. In all systems, bursts can have a variable duration and can have variable intra-burst intervals. This is consistent with a general observation abserved accross all systems: that of the involvement of calcium. VGCCs are involved in generating complex spikes and plateau potentials. Elevation of pre-synaptic calcium is central to short-term facilitation. Elevation of post-synaptic calcium is also essential to the expression of many forms of LTP/LTD.  Defining burst based on a nonlinear readout of calcium means that frequency and number of spikes are both required to establish the boundary between burst and non-burst (Fig. \ref{fig:What}).

\section{Disjunctive and conjunctive burst generation}

\begin{figure*}
    \centering
    \includegraphics{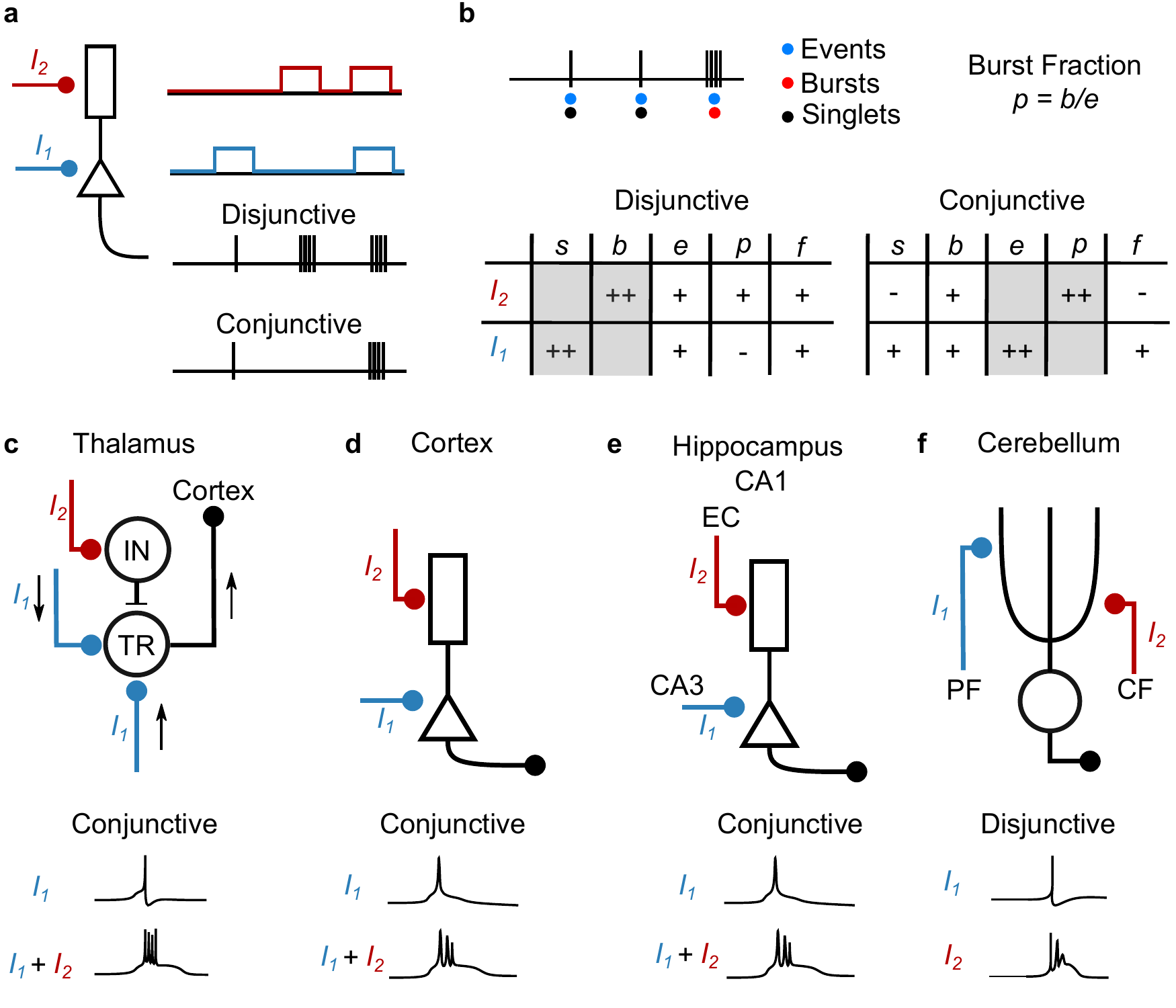}
    \caption{Conjunctive and disjunctive burst generation. \textbf{a} When two input pathways impinge on a cell, bursting arises when the burst inducing pathway is active (disjunctive code) or when both input pathways are active together (conjunctive code). \textbf{b} Separating the spike train into events made of either singlets or bursts and computing the burst fraction by dividing the burst rate by the event rate, we can consider the correlation between singlet rate, burst rate event rate, burst fraction, and firing rate with either inputs. A disjunctive code shows strong correlation between singlet rate and $I_1$ and a strong correlation between burst arte and $I_2$. A conjunctive code shows a strong correlation between event rate and $I_1$ and between burst fraction and $I_2$. \textbf{c-f} illustration pathways giving rise to conjunctive or disjunctive code in the thalamus, cortex, hippocampal CA1 and cerebellum. }
    \label{fig:Conjunctive}
\end{figure*}

Let us consider two fictitious synaptic pathways, pathway `1’ and pathway `2’. Pathway 1 drives action potential firing through leaky integration, with adaptation, refractoriness and feedforwad inhibition to make bursting less likely.  Pathway `2’ is required for burst firing. Now pathway `2’ may drive bursting in two fundamentally distinct ways. It may act autonomously, triggering bursts without input in pathway `1’.  At every time step we may get either an isolated spike from pathway `1’ or a burst from pathway `2’. Alternatively, pathway `2' may drive bursting only when pathway `1’ has generated an action potential. In this mode, at every time step we may get a burst if inputs from both pathway `1’ and pathway `2’ are present. We refer to these two possibilities as disjunctive (1 \emph{or} 2) and conjunctive (1 \emph{and} 2) burst generation. 

Mathematically, we may denote the singlet rate $s$ and burst rate $b$ as functions of two independent inputs $\mathcolor{blue} I_{\mathcolor{blue} 1}$ and $\mathcolor{red} I_{\mathcolor{red} 2}$. In a \textbf{disjunctive} code this means that $\mathcolor{red} b(\mathcolor{red} I_{\mathcolor{red} 2})$ is independent of $\mathcolor{blue} I_{\mathcolor{blue} 1}$ and conversely $\mathcolor{blue}s(\mathcolor{blue} I_{\mathcolor{blue} 1})$ is independent of $\mathcolor{red} I_{\mathcolor{red} 2}$. The rate of either types of events $e = s+b$ will correlate input 1 and with input 2, but more weakly because information from input 2 perturbs information from input 1 and conversely. For a fixed number of spikes per burst $n$, the firing rate $f = s + nb$ also shows a mixed dependence. In a \textbf{conjunctive}, input 1 causes either spikes or bursts such that it is the event rate which is dependent on input 1: $\mathcolor{blue}e(\mathcolor{blue} I_{\mathcolor{blue} 1})$. The burst fraction is controlled by the other input: $\mathcolor{red} p(\mathcolor{red} I_{\mathcolor{red} 2})$. In this code, the burst rate $b = pe = \mathcolor{blue}e(\mathcolor{red}p$ shows a mixed dependence. So does the firing rate ($f=\mathcolor{blue}e(1-(n-1)\mathcolor{red}p)$) and the singlet rate. This structure of correlation is summarized in Fig. \ref{fig:Conjunctive}. The fact that in a conjunctive code, the firing rate remains highly correlated with burst rate can explain why many studies have rejected burst coding in favor of rate coding \cite{Reinagel1999a,Gabbiani1996a,Shinomoto2005a}.


\subsection{Disjunctive dendritic calcium spikes - cerebellum}

Purkinje cells of the cerebellum emit two distinct types of potentials. Simple spikes are short action potentials (1 ms) mediated by sodium and potassium ion channels (Pathway 1). The occurence of these potentials is modulated by inputs from parallel fibers (PF), which target more distal dendrites. Complex spikes are bursts mediated mainly by calcium currents in the dendrites (Pathway 2). These potentials are triggered by input from climbing fibers (CF), which target more proximal parts of the dendritic arborization of these cells. CF input alone is sufficient to trigger complex spikes, while PF inputs are not reported to do so. Dendritic organization has been implicated in regulating the calcium currents underlying complex spikes \cite{Rancz2010a}, but they may not be essential for this disjunctive take place \cite{Davie2008a}. These observations, which are echoed by further observations in vivo \cite{Herzfeld2015a}, point to a disjunctive burst code in the cerebellum.

\subsection{Conjunctive bursting – hippocampus}

 Pyramidal cells of the CA1 region have a prominent apical dendrite and a number of basal as well as radial oblique dendrites, segregating input from different pathways. Action potentials can be triggered by input onto basal and radial oblique, but input onto apical tuft undergo much electrotonic attenuation and are not thought to contribute as strongly to action potential generation. Consistently, inputs from CA3, which targets radial oblique dendrites can trigger single action potentials but rarely bursts \cite{Pouille2001a,Takahashi2009a}. These inputs can be seen as an example of pathway `1’, from the previous section. Inputs onto the apical dendrite may be seen as pathway `2’ as they can produce plateau potentials \cite{Takahashi2009a}. Unlike in the cerebellum where a single stimulation was sufficient to trigger a complex spike, very strong stimulation of the perforant path targetting the apical dendrites is required to generate a plateau potential. Concomittant but relatively weak stimulation of afferents from CA3 and from entorhinal cortex are sufficient to elicit these plateau potentials \cite{Takahashi2009a}. These in vitro studies suggests a conjunctive code, which has seen further experimental support in vivo  \cite{Bittner2015a}.  Such conjunctive burst generation is expected to arise for well separated pathways for a wide range of dendritic spike shapes and strength of the back-propagating action potential \cite{Friedenberger2022a}. In vitro studies have further supported the role of the back-propagation of somatically generated action potentials, distal NMDA-spikes and the large amplitudes of distal post-synaptic potentials in shaping input integration in these neurons \cite{Stuart1994a,Magee1999b,Magee1999a,Grienberger2014a}.

Principal cells from other parts of the hippocampus were also observed to generate bursts of action potentials. This includes cells from the dentate gyrus \cite{Kowalski2016a} and cells from the CA3 \cite{Hablitz1981a,Balind2019a}. Granule cells from dentate gyrus can produce dendritic spikes with durations around 40 ms with a strong attenuation of the backpropagating action potential \cite{Krueppel2011a}, but little is known about whether these mechanisms are associated with the generation of high-frequency bursts in these cells. Cells from the CA3 have calcium-based dendritic spikes that give rise to bursts \cite{Balind2019a}. The back-propagating action potential appears to help the generation of these calcium spikes \cite{Mago2021a},but more evidence is required to establish if the CA3 pyramidal cells are using a conjunctive code. In these cells, both recurrent connections and input from dentate gyrus are more likely to trigger single action potentials (pathway `1’), while input from entorhinal cortex could be modulating the burst fraction. Evidence from in vivo recordings would help establish the conjunctive nature of CA3 bursting with more certainty. 

\subsection{Conjunctive bursting – cortex}

Cortex has many types of principal cells across its layered structure \cite{Tasic2018a}. Deep layer, thick tufted neurons with projection to the pyramidal tract neurons can generate burst of action potentials mainly when input to the apical dendrite is combined with an action potential \cite{Larkum1999a,Larkum2001a,Larkum2004a}. Similar to cells from CA1, inputs to basal and radial oblique dendrites receive recurrent and thalamic connections (Pathway `1’ \cite{Yaeger2022a}) whose ability to generate action potentials at a high frequency is counteracted by potent feedforward inhibition \cite{Sermet2019a}. These mechanisms suggest a conjunctive code whereby apical inputs AND input basal/radial oblique input are required to generate bursts. Xu et al 2012 \cite{Xu2012a} have confirmed the conjunctive nature of burst coding in vivo by showing that both calcium plateau potentials in somatosensory cortex arise only in the combination of sensory input and feedback from motor cortex. Cortex appears to use a conjunctive code where feedforward inputs from sensory thalamus or lower order cortical areas (pathway `1’) must combine with feedback inputs from higher-order thalamus or higher-order cortical areas (pathway `2’) to generate bursts of action potentials. 

Bursts of action potentials is also known to occur in vivo in pyramidal cells of the layer 2-3 \cite{deKock2008a,deKock2021a,Wang2020a} and in spiny stellate cells \cite{Brecht2002a}. The superfucial layers tend to produce more burst than the deeper ones \cite{Shinomoto2009a}.  The absence of either potent calcium spikes (for L2-3 cells \cite{Larkum2007a}) or apical dendrites (for spiny stellate cells) would indicates that these bursts do not arise from the same mechanism as in deep layer pyramidal cells. NMDA-spikes and dendritic sodium-ion channels can control burstiness and have been observed in these cells \cite{Brumberg2000a,Palmer2014a,Smith2013a,Friedenberger2022a}.


\subsection{After-hyperpolarization rebound – thalamus}

Thalamic relay neurons are known to produce bursts of action potentials in vivo, regularly under anesthesia and drowsiness and sporadically under awake conditions \cite{Swadlow2001a}.  Electrophysiological investigations in slices have identified a potent bursting mechanism relying on T-type voltage-gated calcium channels \cite{Deschenes1984a,Huguenard1992a,Wang1991a}. In those cells, a normal somatic current injection produces regular firing, but the same current injection preceded by a long (100 ms) period of hyperpolarization produces a burst of action potentials. Bursts elicited in this manner requiring both a sustained inhibition and then an excitation, hinting at a form of conjunctive bursting, but one with an implicit delay between the two inputs. 
A study of tracking thalamic relay cells firing patterns in vivo across behavioral states \cite{Urbain2015a} has shown independent modulation of firing rate and burst fraction. The study noted, however, that bursts were not preceded by periods of hyperpolarization, suggesting that bursts did not arise from after-hyperpolarization rebound. Another study made a more direct test to the role of hyperpolarization in controling bursting \cite{Borden2022a}. With optegenetic hyperpolarization of thalamic neurons, the firing rate changes only weakly while there is a large increase in the burst fraction. These result show that controling the hyperpolarization controls the burst fraction, without affecting the firing rate. Yet another in vivo study brought further support to this view \cite{Born2021a}. Tracking firing patterns while optogenetically supressing corticothalamic activity, they found that the firing rate decreases, consistent with a net excitatory effect of corticothalamic afferents. Surprisingly, this was accompanied by a an increase in the burst fraction. Since visual afferents providing the main source of excitation to these cells shows a co-modulation of bursts and single spikes \cite{Reinagel2000a}, these observations are consistent with a conjunctive code where net inhibition onto the thalamic cells controls the burst fraction while net excitation controls the firing rate.

\section{Burst-dependent synaptic plasticity}

Long term potentiation of synaptic efficacy (LTP) provides a compelling cellular basis for learning and memory \cite{Nicoll2017a}. Reliably triggered 50 years ago by intense stimulation of hippocampal pathways \cite{Bliss1973a}, researchers have now established some of the core elements necessary for its expression. LTP requires NMDARs \cite{Collingridge1983a,Morris1982a}, elevated post-synaptic calcium \cite{Lynch1983a}  as well a local glutamate release \cite{Lledo1998a}. These elements suggest a model whereby LTP arises from elevated post-synaptic calcium taking place relatively soon after or before some amount of glutamate has been released at a synapse. 

Multiple chemical pathways depend on this elevation of post-synaptic calcium \cite{Bhalla1999a,Maki2020a}. For calcium to be elevated at the synapse, it may arise from NMDAR-mediated currents, local release from calcium stores or VGCC. Whereas the first two sources are controlled by local synaptic inputs, VGCCs make calcium entry dependent on local voltage and thus possibly remote agents can control plasticity if they can control the voltage across the dendritic tree. Because bursts, particularly when accompanied with a sustained depolarization, would elevate the local membrane potential, reaching into the majority of dendrites \cite{Lisman2005a,Stuart1994a,Nevian2006a}, these events are uniquely positioned to control LTP expression.

This view focuses on the role of post-synaptic bursting in regulating LTP/LTD, a view that is consistent with previous literature focusing on the role of the relative timing between pre-and post-synaptic spikes. When formulating learning rules based on relative timing on pre- and post-synaptic side of the synapse, many studies have considered the role of the post-synaptic firing patterns in regulating system-level plasticity. This model is essential to explain the dependence on repetition frequency observed in protocols of spike-timing dependent plasticity (STDP) \cite{Sjostrom2001a,Froemke2006a,Pfister2006a}. Multiple modeling studies have indicated that including burst-dependence in learning rules allows for the emergence of various types of selectivities. This includes the selectivity of ON- OFF- pathways in the developping retinogeniculate connection \cite{Butts2007a,Gjorgjieva2009a}, orientation selectivity in the developping thalamoctortical afferents \cite{Toyoizumi2005a,Pfister2006a,Ren2022a} as well as finer patterns of recurrent connectivity within cortex \cite{Clopath2010a}. 

The focus on post-synaptic bursting has conceptual implications that require a departure from Hebbian plasticity. Donald Hebb has speculated that learning takes place when a pathway taking part in firing a post-synaptic cell would be potentiated. This picture is entirely consistent with the simplified picture of LTP expression above, as long as a (feedforward / pathway 1) pathway can take part in creating a burst, this pathway will be potentiated. But Hebb’s theory needs to be revised when bursting is not triggered by the same pathway undergoing the potentiation. This is striking for a disjunctive burst code where a given pathway instructs the plasticity of another pathway as in the cerebellum \cite{Marr1969a,Albus1971a,Ito1982a}. A process that is better called 'instructive plasticity' \cite{Grienberger2021a} or 'Marr-Albus-Ito Plasticity'. For a conjunctive code, the departure from Hebb is more subtle. Bursting, and therefore LTP, can be engaged when two pathways come together. This form of plasticity is partially Hebbian because of the need of co-activation between pathway 1 and the post-synaptic neuron. But in a strict interpretation, it is non-Hebbian because bursting is caused by the presence of another, instructive, input, itself not necessarily undergoing plasticity.

In vitro recordings have confirmed such a central role of post-synaptic bursting in the expression of LTP. In physiological calcium, pairing an action potential with a post-synaptic spike shows LTD while pairing with a post-synaptic burst shows LTP \cite{Inglebert2020a}. The timing between the pre-synaptic spike and the post-synaptic burst need not be precise as the two events can be separated by up to 1 second \cite{Bittner2017a}. LTD can also arise from pairing pre-synaptic activity with post-synaptic bursts separated by half a second \cite{Milstein2021a}. Crucially, such behavioral timescale plasticity (BTSP) \cite{Grienberger2021a} has been shown to induce or tweak place selectivity in hippocampus CA1 \cite{Bittner2017a,Milstein2021a}. Further work will likely establish the role played by BTSP in cortex \cite{Aru2022a}.

Plasticity in the cerebellum also shows burst dependence, but here the association of an eligibility trace with a post-synaptic burst leads to LTD while pairing with a singlet leads to LTP \cite{Coesmans2004a,Lev2002a,Jorntell2006a,Bouvier2016a}. These plasticity rule show symmetric depenence on the ordering of pre- and post-synaptic spikes \cite{Suvrathan2016a}, much like in similar burst-dependent learning rules in the electric sense of electric fish \cite{HarveyGirard2010a,HarveyGirard2013a,Bol2011a,Mejias2013a}.

\section{Differential transmission of firing patterns}

\begin{figure*}
    \centering
    \includegraphics{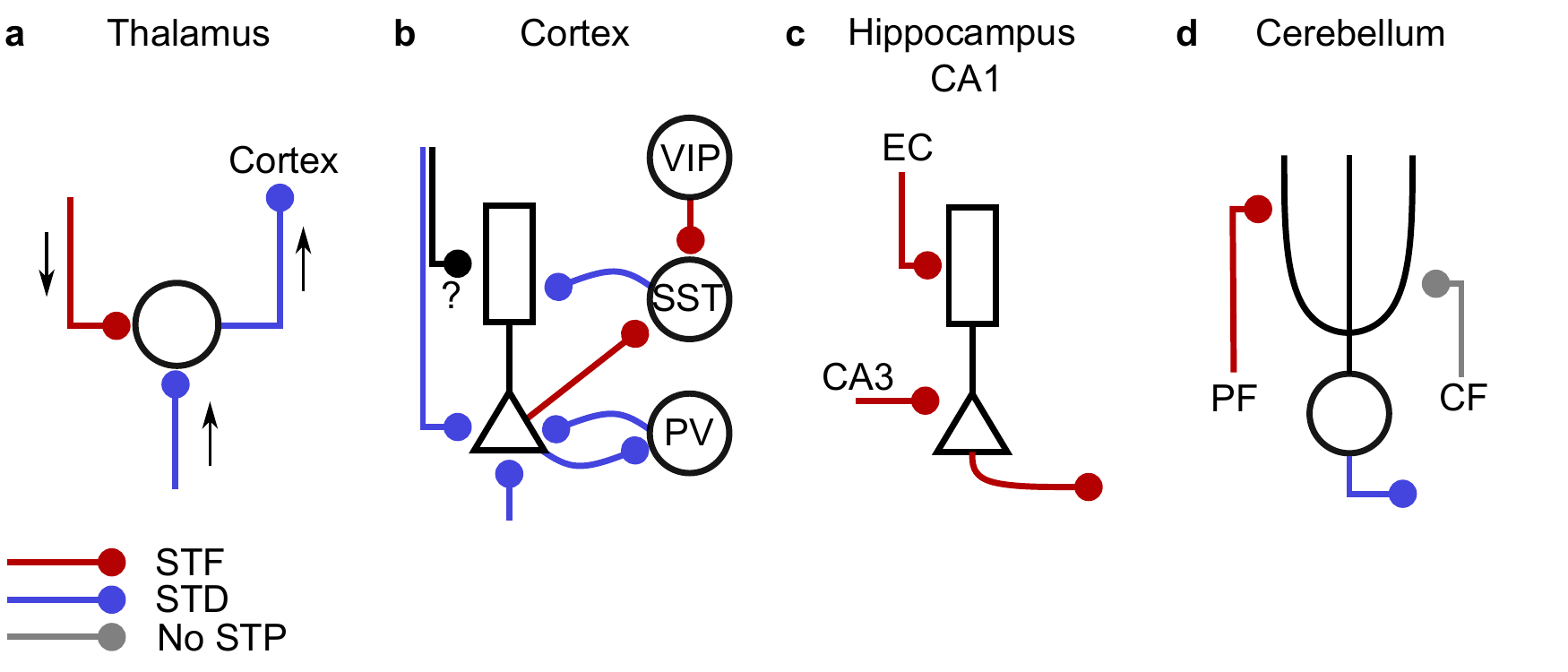}
    \caption{Patterns of short-term plasticity in four principal cells. \textbf{a} Sensory thalamus receive STD from the retina \cite{Granseth2002a}, sends STD to cortex \cite{Swadlow2001a,Cruikshank2010a} and receives STF from cortex \cite{Granseth2002a,Cruikshank2010a,Jackman2016a}.\textbf{b} Pyramidal cells of sensory cortex receive STD from thalamocortical afferents \cite{Swadlow2001a,Cruikshank2010a} and from motor cortex \cite{Lee2013a}. The Interconnections with PV-positive cells is STD \cite{Campagnola2022a} and connections onto SST cells is STF both from glutamatergic cells and from gabaergic VIP-positive cells \cite{Campagnola2002a}. As some top-down projections target the soma and other top-down projections target the apical dendrites \cite{Geng2022a}, we have assumed that STP experiments only revealed proximal connections. \textbf{c} Inputs from both entorhinal cortex (EC) and CA3, onto CA1 pyramidal cells, are facilitation \cite{Jackman2016a}. Outputs onto the subiculum are facilitation \cite{Xu2012b}. \textbf{d} Climbing fiber (CF) inputs onto Purkinje cells shows no STP in 1~mM calcium \cite{Foster2002a}. Parallel fiber inputs onto Purkinje cells show STF \cite{Dittman2000a}. Outputs from the cerebellum to the deep cerebellar nuclei (DCN) undergo STD \cite{Telgkamp2002a,Jackman2016a}.} 
    \label{fig:STP}
\end{figure*}


To separate bursts from single spikes post-synaptically, transmission must be frequency dependent. No dendritic computation can reliably separate a burst in one afferent from synchronous single spikes across multiple afferents. Instead, we must consider frequency-dependent transmission that is specific to one afferent.  Following a similar reasoning, Francis Crick had hypothesized such properties of synapses, which he called von der Marlsburg synapses \cite{Crick1984b}. Today, the existence of STP is well established. Yet to align specifically with a function of burst transmission, synaptic dynamics should match the main features of burst generation. Much of the early research on STP have focused on facilitation over timescales slower 50 Hz \cite{Salin1996a,Zucker2002a,Varela1997a,Tsodyks1998a,Toth2000a}. More recent experiments have confirmed the existence of STP that is triggered specifically by burst-like events having a frequency closer to 100 Hz \cite{Chamberland2018a,Lefort2009a,Jackman2016a,Campagnola2022a,Vandael2020a}, with a sensitivity to lengthening of the spike duration \cite{Geiger2000a} but not to sustained depolarization \cite{Apostolides2016a}.

In this way, a mixture of spikes an bursts can be demixed by STP. When the same axon projects with STD at one target and STF at another, synaptic projections can communicate independent information - two streams of information - to different post-synaptic targets \cite{Naud2018a}. What logic is there to STP in synaptic projections? The different types of projections may reflect a different functional role on the post-synaptic target, such as a driver vs a modulator role \cite{Sherman2001a,Sherman2001b,Sherman2012a}. Alternatively, the different types of projections may reflect different filtering operations on the pre-synaptic spike train \cite{Izhikevich2003c,Fortune2001a}, allowing for routing of information \cite{Kording2000a,Naud2018a,Payeur2019a,Payeur2021a}. In the latter, STP inherits the semantic of bursts and singlets: applying STD on a set of spike trains $S$ extracts the event rate $e= STD[S]$ and applying STF extracts the burst rate $b = STF[S]$. In a conjunctive code, this means that STD connections communicate information from pathway 1 and that STF projections communicate information from a conjunction of pathway 1 and 2.

\subsection{Cortex} 

A recent extensive study in cortex \cite{Campagnola2022a} has surveyed more than 20 000 synaptic pairs across 1502 animal slices.  The study also reported over 2500 synaptic pairs in human tissue. Campagnola et al. (2022) \cite{Campagnola2022a} found that, most connections showed some level of STD over a wide range of stimulation frequencies. Separating cells according to transcriptomic families, they found that connections to and from PV-positive cells tend to show depression. These observations echo previous studies that have shown a tendency for depression for local connections onto fast-spiking or basket-type cells \cite{Reyes1998a,Markram1998a,Tsodyks1998a}. STD is also observed when PV-positive cells connect to their post-synaptic partner \cite{Galarreta1998a}. 

Many other synapses show STD, and multiple type of temporal dependencies are observed. At some synapses, but present in both human and mice cortex, a strong depression kicks in at frequencies of at least 100 Hz. A burst attempting to cross this type of synapse would communicate only the first few spikes in the burst – specifically those spikes that are triggered by pathway `1’ in a conjunctive code \ref{fig:Conjunctive}. The depression takes between half a second and a second to recover, so that isolated spikes that follow a burst would also be attenuated. This type of frequency-dependent depression is observed in L4 pyramidal to PV-positive cell connections \cite{Campagnola2022a}. 

Similarly, many synapses show STF, and multple types of temporal dependencies are observed. At some synapses, a frequency-dependence with a cutoff at 50 or 100 Hz is observed \cite{Lefort2009a,Campagnola2022a}. This facilitation tends to cumulate over multiple high-frequency spikes. Campagnola et al. (2022) \cite{Campagnola2022a} found that among local cortical connections and separating according to transcriptomic families, it is those connections onto and from SST- and VIP-positive cells that display STF. This again echoes previous studies based on morphology or firing patterns of inhibitory cells that have indicated a tendency for connections onto Martinotti cells (SST-positive) to display potent facilitation \cite{Berger2010a,Tsodyks1998a,Reyes1998a}. 

The occurrence of frequency-dependent STF and STD suggests that different synapses will communicate either selectively bursts (STF) or both bursts and single spikes with equal rate (aka the event rate). Combining these synaptic properties in cellular connectivity motifs (i.e. feedforward inhibition) can also implement a selectivity to the fraction of bursts \cite{Naud2018a}. Cortical networks are thus able to separate information from pathway ‘1’ and ‘2’ in a conjunctive code. Theoretical studies have further established that when feedback inhibition is separated in two groups, one receiving STF and the other STD, network receiving two streams of input in a conjunctive burst code show better balance and better information transmission \cite{Naud2018a,Keijser2020a,Vercruysse2021a}.

Is this polarization of local connections in terms of STD and STF also reflected in a polarization of impinging pathways? In sensory cortices, input are broadly separated into feedforward (e.g. sensory afferents) and feedback (e.g. cortico-cortical, high order thalamus) connections. Feedforward inputs to the cortex target principal neurons with STD and PV-positive cells providing feedforward inhibition also with STD \cite{Gibson1999a,Gil1999a,Beierlein2003a,Gabernet2005a,Jackman2016a}. These results, however, were obtained in young animals, such that age may reduce the potency of STD \cite{Frick2007a,Oswald2008a}.  Descending connections, on the other hand, come in two types \cite{Geng2022a}. The apical type targets apical dendrites of pyramidal cells, SST and VIP cells, and the proximal type targets both apical and proximal dendrites as well as PV positive cells. Connections onto SST and VIP cells tend to express STF \cite{Sylwestrak2012a,Campagnola2022a}, which would support a polarization of STP along ascending and descending pathways, but this polarization remains mainly an assumption of cortical information processing theories \cite{Naud2018a,Payeur2021a,Greedy2022a}.   

\subsection{Hippocampus}

In the hippocampus, and focusing on CA1 pyramidal cells, the perforant path (PP) and the schaffer collateral (SC) are the two main pathways (Fig. \ref{fig:STP}c). PP inputs from entorhinal cortex \cite{Witter2006a} targets mainly apical dendrites and NPY-positive cells, with a smaller fraction of input going to SST-positive cells \cite{Kajiwara2008a,Milstein2015a}. These inputs show weak STF with a frequency dependence that is controlled by the presence of inhibition \cite{Milstein2015a}.  SC inputs from CA3 target mainly apical oblique dendrites, PV-, SST- as well as NPY-positive cells. These connections all show STF with frequency dependence \cite{Milstein2015a,Wierenga2005a}. This suggests an preponderance of STF for inputs onto CA1, a trend that extends to other parts of the hippocampus \cite{Salin1996a,Toth2000a,Rossbroich2021a}. CA1 pyramidal neurons then project to multiple targets, with many connections going to the subiculum. These connections also display STF \cite{Xu2012b}. This picture paints a picture dominated by facilitation into and out of the hippocampal CA1. 

\subsection{Cerebellum}

Purkinje cells are the most numerous cells of the cerebellum. They receive inputs from granule cells in parallel fibers (PF) and from the inferior olive in climbing fibers (CF) (Fig. \ref{fig:STP}d). While motor related commands (and reward-related information) are carried by PF, the CF carry motor error (and reward-related) information \cite{Ito2008b,Kostadinov2022a}. PF inputs show STF \cite{Dittman2000a}, while CF inputs show an absence of potent forms of short-term plasticity \cite{Foster2002a}. 

\subsection{Thalamus}

Feedforward infomation arriving from the senses to sensory thalamus shows STD \cite{Granseth2002a,Reichova2004a}. When this information continues to sensory cortices, thalamocortical afferents, whether to excitatory or inhibitory neurons, also show STD \cite{Gibson1999a,Gil1999a,Beierlein2002a}. When the cortex sends feedback information to sensory thalamus, it does so with STF projections
\cite{Granseth2002a,Reichova2004a,Cruikshank2010a,Jackman2016a}. But when cortex sends to higher-order thalamus, the information could be considered as part of a feedforward stream. These afferents display STD (PoM \cite{Mease2016a})).

\subsection{What logic}

What logic arises from these connectivity patterns? Inputs to a principal cell are both STD and STF in the thalamus, more STD in the cortex and strictly STF in the hippocampus. Outputs from principal cells are strictly STD in the cerebellum and thalamus, but STF in the hippocampus. The idea that STD is used to communicated feedforward information is supported by the thalmus and to some extent the cortex, but the cerebellum appears as an exception since error-carrying CF shows no potent plasticity. Another idea is that STF inputs are used to control LTP. This rule applies to the cortex as SST-positive cells have been strongly implicated in the control of bursting and therefore of LTP \cite{Urban2016a,Artinian2018a,Chen2015a,Doron2019a}, in the hippocampus as both EC and CA3 inputs are required for LTP \cite{Takahashi2009a} but also in the cerebellum as PF inputs are associated with LTP \cite{Coesmans2004a,Lev2002a}. Clearly, however, much is missing from the picture of connectivity types in all of these systems.

 \section{Algorithmic requirements for burst coding}

Why would it be beneficial for the nervous system to represent information using two distinct types of activity? What types of algorithms would be difficult to implement without this ternary code? We have described how properties of neurons and synapses in line with the burst coding hypothesis allow neurons to communicate, represent and exploit two types of syllables, but what algorithm would benefit from such a separation? We have argued that a three-syllable code allows to represent, process and transmit two types of information with one associated with elevated levels of intracellular calcium. We now review the place of such multiplexing in the theory literature.

One category of algorithm that necessitate two streams of information are learning algorithms. For the most part, supervised learning proceeds by representing and communicating two types of information: information about the inputs to the network and information about the relative error with respect to a target response. thms allow synapses to change in such a way as to optimize a global objective. The backpropagation of error is one thm. It is not plausible to implement its exact formulation in the brain, but it is possible to enact approximations \cite{Lillicrap2020a}. Dendrite-dependent bursting combined with burst-dependent plasticity and polarization of short-term plasticity have been used in a computational model to approximate the backpropagation of error algorithm \cite{Payeur2021a,Francioni2022a,Greedy2022a}. 

As for reinforcement learning, with its powerful instances that uses deep neural networks, these algorithms rely on supervised learning techniques. Here, a reward prediction error is used to guide the output of a neural network, but the neural network then solves the credit assignment problem using backpropagation-like algorithms \cite{Hassabis2017a}. Therefore, deep reinforcement learning algorithms would benefit from burst coding for the same reason that supervised learning does. 

Within the category of learning algorithms, there is also unsupervised learning. A powerful type of unsupervised approach called the 'wake-sleep algorithm' can be used to train various model structures (e.g. Helmoltz machines). These algorithms proceed by exchanging two types of information in a network \cite{Hinton1995a,Vertes2018a}, one information about observations and one about predicted observations. Since the algorithm proceeds by comparing these representations, both must be represented approximately at the same time, an algorithmic demand that can be implemented using burst multiplexing. Other unsupervised algorithms proceed by requiring particular statistical patterns onto higher-order representations. These sometimes exploit the backpropagation of error algorithm \cite{Zhuang2019a}, and at other times succeed in communicating a single type of information to other neurons but require plasticity to depended on two types of activity \cite{Illing2021a,Halvagal2022a}. In every case, bursting appears as a natural candidate for the implementation of aspects of these algorithms.

In machine learning, learning-related signals can be processed in many different ways. It is not necessary for these signals to follow a simple flow pattern such as a top-down backpropagation across a hierarchical network. Researchers have highlighted the importance of memorizing or projecting in space/time such learning-related signals \cite{Jaderberg2017a,Neftci2019a}.

Another category of algorithms closely linked to learning are those of attention. Attention selectively coordinates the enhancement or suppression of task-relevant sensory representations using top-down feedback. Theories are typically concerned with attention's role in perception and view attention as biasing the competition between competing stimuli \cite{Reynolds2009a, Desimone1995a} or aiding in the binding of features within the visual scene \cite{Treisman1980a}. When implemented in theoretical models, these attention signals take the form of an additive or multiplicative gain modulation of neurons representing the feature of interest. On the other hand, there are theories concerned with attention's role in learning and view attention as a signal that gates plasticity \cite{Roelfsema2005a, Roelfsema2018a}. With the expectation of attention within the prediction coding framework \cite{Feldman2010a}, these two perspectives are traditionally treated separately. Within the burst coding framework, multiplexing is well suited to facilitate the coordination of two independent streams of information up and down the cortical hierarchy, where top-down "attention-like" feedback signals targeting dendrites can be a source of both gain modulation and plasticity gating. Therefore, burst coding is well suited to link these two roles for attention.

\section{Bursting in vivo}

\begin{table}
\tabcolsep7.5pt
\caption{\label{tab:anesth} Stationary burstiness Under Anesthesia. $^{\rm a}$: ratio of bursty intervals to all intervals.; $^{\rm b}$ bursts require a silent period before (100 ms).}
\label{tab1}
\begin{center}
\begin{tabular}{@{}l|l|c|c|c@{}}
 & & $\Delta$ &BF\\
Area& Type & (ms) & (\%) & Ref.\\
\hline
\textbf{S-CTX}&
Rat, S1  &10 &1$\pm$1$^{\rm a}$ & \cite{deKock2008a}\\
& L2-3 &&& \\
&Rat, S1 &10&17$\pm$5$^{\rm a}$ &\cite{deKock2008a}\\
&  L5B  && 5& \\
\hline
\textbf{HP}&Rat, CA1 & $\sim$20 & 96.3$\pm$0.8$^{\rm a}$&\cite{Kowalski2016a}\\
& &&& \\
&Rat, CA3 & $\sim$20  & 86.5$\pm$2.4$^{\rm a}$ &\cite{Kowalski2016a}\\
 & &&&\\
\hline
\textbf{THL}&
Cat, LGN   &4 & 13-25$^{\rm b}$  & \cite{Lesica2004a}\\

\end{tabular}
\end{center}
\end{table}

\begin{table}
\tabcolsep7.5pt
\caption{\label{tab:awake} Stationary Burstiness Quiet Awake / freely moving.  $^{\rm d}$ plateau potentials identified based on presence of sustained depolarization. $^{\rm e}$ ratio of the number burst to total number of action potentials.$^{\rm f}$ ratio of the number burst to total number of events (bursts or singlets).}
\label{tab1}
\begin{center}
\begin{tabular}{@{}l|l|c|c|c@{}}
 & & $\Delta$  &BF&\\
Area& Type & (ms) &  (\%)  &  Ref.\\
\hline
\textbf{S-CTX}&Monkey,V1, & 6 & 35$\pm$6$^{\rm a}$  &\cite{Onorato2020a}\\
& L2-4  &&&\\
&Rat, S1,  &10&15$\pm$3$^{\rm a}$  &\cite{deKock2008a}\\
& L2-3 &&&\\
&Rat, S1 &10&17$\pm$5$^{\rm a}$  &\cite{deKock2008a}\\
& L5B  &&&\\
&Rat, S1 &15& $\sim$ 12$^{\rm f}$  &\cite{Naud2022a}\\
& L5  &&&\\
\hline
\textbf{F-CTX}&Monkey, ACC/PFC & 5 & 30-40$^{\rm e}$  &\cite{Womelsdorf2014b}\\
\hline
\textbf{HP}&
Mouse, CA1   &-$^{\rm d}$ & 3.8$\pm 8^{\rm e}$  & \cite{Bittner2015a}\\
& Rat, CA1   &50 & ~ 40$^{\rm a}$  & \cite{Epsztein2011a}\\
& Rat, CA1   &10  &  ~ 29-46$^{\rm a}$ & \cite{Sanders2019a}\\
& Rat, CA3   &10  &  ~ 29-37$^{\rm a}$ & \cite{Sanders2019a}\\
& Mouse, CA1 & 15 & 20$^{\rm a}$ & \cite{Tanaka2018a}\\
& Mouse, DG & - & 65$\pm $ 22 & \cite{Pernia2014a}\\
\hline
\textbf{THL}&
Mouse,dLGN   &4 & 5$^{\rm b}$  & \cite{Born2021a}\\
&Rabbit, dLGN &  & 5-40 $^{\rm b}$& \cite{Bezdudnaya2006a}\\
&Mouse,VPM   &10 & 15-20 $^{\rm a}$ & \cite{Urbain2015a}\\
&Mouse,Pom   &10 & 10-15 $^{\rm a}$ & \cite{Urbain2015a}\\

\end{tabular}
\end{center}
\end{table}

Are bursts being spontaneously generated in the awake and behaving animal? and if so, are these events only occurring at specific times? Either using patch clamp or extracellular recording techniques, many \textit{in vivo} studies have observed firing patterns in the behaving animals. While the answer to the first question is a clear yes, the second question is more difficult to answer. Bursts occur spontaneously at a low rate in many brain areas, both under anesthesia and in awake animals (Table  \ref{tab:anesth} and \ref{tab:awake}). The fact that they occur spontaneously in all conditions also means we cannot equate their occurrence with an input or a coding feature, but instead we must resort to a comparison of the relative amount of burst that have occurred. 

Unfortunately, many technical aspects renders cross-study comparisons particularly difficult. The first aspect is that bursting is strongly correlated with firing rate (Fig. \ref{fig:Conjunctive}). While a condition or state may show more bursting, when this is explained by a difference in firing rate means it is not about bursting but about firing rate. For this reason, researchers tend to measure the burstiness by normalizing with some measure of firing rate. There are multiple ways of computing such burst fractions, and this diversity confuses comparisons. For instance, one may divide the number of bursts by the total number of AP or divide the number of short ISIs by the total number of APs. The latter will always give larger values than the former. Similarly, more stringent criteria for bursting (ISI cutoff or requiring a period of silence before a short ISI) will necessarily lead to lower burst fractions. Other factors include the recording methods, while in vivo patch-clamp allow one to measure the presence of a sustained depolarization during bursting, these techniques are very challenging and are not the majority. Extracellular recordings are more convenient but introduce potential artefacts due to spike sorting. Absent or incomplete spike sorting will tend to increase the firing rate and burst rate. Furthermore, spike sorting tends to sort-out spikes with altered spike shape, thus sorting out later spikes in a burst. With these technicalities in mind, we have compiled experimentally reported burst fractions in Tables \ref{tab:anesth} and \ref{tab:awake}. 

The anesthetized state has been known to alter the burstiness of neurons. This was highlighted by studies in sensory thalamus where both anesthesia and sleep are associated with higher degrees of burstiness \cite{Ramcharan2000a,Mukherjee1995a}. This tendency is further reflected in the separation between inattentive state with higher burstiness and attentive states with the lowest relative proportion of bursts \cite{Swadlow2001a} (but see \cite{Urbain2015a}).  Does this relationship between arousal and bursting extend also to other areas? In the hippocampus, anesthesia is also associated with a state of predominantly burst firing (Tab. \ref{tab:anesth}), which recovers a sparse mixture in awake states (Tab. \ref{tab:awake}). Cortex shows a mixed yet contrarian picture: while L5 pyramidal neurons show no dependence on anesthesia, L2-3 pyramidal neurons stop firing bursts altogether in anesthetized states \cite{deKock2008a}.

In awake states, the burst fraction appears fairly uniform across areas and species. It is, for the most part, sitting between 15 and 40\%. Similarly, the average number of spikes per bursts typically sits between 2 and 3. Such states of sparse bursting is optimal for communicating two streams of information \cite{Naud2018a}. Within this range, there are variations across areas. Within the hippocampus, granule cells have the highest proportion of bursts; and within the cortex burstiness changes across lamina with a minimum in L5 and higher values in L2-3 and L6 \cite{Shinomoto2005a,Senzai2019a}. Variations across areas would indicate a greated propensity for bursts in visual and prefrontal areas, as compared wiwth motor areas \cite{Shinomoto2009a}.

\subsection{Attention states}

Under states of visual attention, the saliency of locations (spatial attention) and features (feature-based attention) in the visual field are selectively enhanced or suppressed in order to increase perceptual performance. Across the visual cortex, modulations of the local field potential (LFP) \cite{Fries2001a, Chalk2010a, Khayat2010a} and noise correlations are observed at the population level \cite{Cohen2009a, Ni2018a, Ruff2016a}, while individual neurons show stimulus selective increases in firing rate \cite{Moran1985a, Spitzer1988a, McAdams1999a, Treue1999a} and decreases in trial-to-trial variability\cite{Cohen2009a, Mitchell2007a, Mitchell2009a}. As bursts can be a source of neural gain and variability, modulations in bursting are expected to influence all of these measures. However, the relationship between bursting and these measures will be dependent on the underlying circuit motifs and burst generation process. Therefore, future modeling work will be needed to determine if these observed changes in firing rate and variability can be explained by modulations in bursting or if they are independent.

Direct measures of bursting modulation with attention have been observed across the visual cortex, in higher and lower areas of the ventral stream (V4 and V1) \cite{Anderson2013a, Huang2019a} and in the medial superior temporal (MST) area of the dorsal stream \cite{Xue2016a}. When spatial attention is directed onto a neuron's receptive field, there is an increase in firing rate and a reduction in burstiness upon evoked activity, as measured using an ISI-based metric (fraction of ISIs < 4 ms) and an autocorrelation-based metric. However, this reduction in burstiness appears to be area, cell-type, and context-specific. First, the reported degree of burst modulation in V1 was less than what was observed in V4, in line with the known polarity in firing rate changes \cite{Buffalo2009a, MartinezTrujillo2022a}. Second, when separating neurons into cell-type specific classes based on the extracellular spike duration, the reduction in burstiness was only significant for broad spiking pyramidal neurons at low firing rates (<20 Hz), and did not occur for narrow spiking inhibitory neurons \cite{Anderson2013a}. Lastly, only neurons whose firing rates increased with task difficulty showed a reduction in burstiness, while those with firing rates suppressed by task difficulty did not show a significant modulation in burstiness \cite{Huang2019a}. The reduction in burstiness also appears to only occur for spatial and not feature-based attention \cite{Xue2016a}, suggesting that spatial and feature-based attention may rely on different mechanisms.

In contrast to the visual areas, attention-dependent increases in bursting have been observed in anterior cingulate and prefrontal cortex (PFC/ACC) \cite{Voloh2018a}. Following the cue during a covert attention task, burst fraction increases for both putative inhibitory and excitatory neurons, as determined by extracellular spike shape. However, the magnitude of the increase was cell-type specific, with putative inhibitory neurons showing a larger increase. Both cell-types also showed differences in their relation to LFP  frequency bands. Within the theta frequency band burst firing coincided with stronger theta more than non-burst firing for both cell types, but was stronger for inhibitory cells. Within the beta frequency band excitatory neurons were associated with strong beta power at phases preceding the preferred phase. Suggesting that the bursting of inhibitory and excitatory neurons in the PFC/ACC may play a role in the synchronization of local population activity during attention states.

“Attention-like” changes in bursting have also been observed in the primary somatosensory cortex of rodents. That is, changes in bursting which correlate with a shift in perceptual performance. During a whisker deflection detection task, activity within the apical dendrites of L5 pyramidal neurons is causally linked to the perceptual detection threshold \cite{Takahashi2016a}. Increases in calcium activity within the dendrites and bursting at the soma were positively correlated with detection probability for a subset of neurons, suggesting that conjunctive bursting influences perception. Similarly, for a microstimulation detection task, perceptual detection depended on the irregularity of stimulation, indicating that bursting makes representations more salient \cite{Doron2014a, Doron2019a}. Learning to respond to the microstimulation task appears to occur in two stages \cite{Naud2022a}. First, a fraction of neurons became selective to the stimulus by slightly increasing their firing rates. During the second stage, the formed representations are sharpened, with only the selective neurons showing an increase in firing rate. This is due to a temporal alignment of burst modulation with the previously learned representations. Taken together, these results suggest that an “attention-like” top-down signal modulating bursting is able to selectively sharpen previously learned task-relevant representations in order to increase perceptual performance.

\subsection{Learning}

A learning-related role of bursting has been studied in the hippocampus in the context of place-field formation. Recordings in behaving mice have indicated that place field are more likely to arise in cells with a propensity to produce bursts \cite{Epsztein2011a}. A correlation between the spontaneous occurrence of plateau potentials and place field formation has been noted \cite{Bittner2015a}. Testing for a causal relationship, Bittner et al. (2017) have found that the artificial induction of a few plateau potentials is sufficient to give rise to a new place field close to the location occupied by the animal at the time of the induction \cite{Bittner2017a,Zhao2022a}.  In vivo bursting in these cells requires inputs from the entorhinal cortex (EC), CA3 and is helped by disinhibition from SST-positive \cite{Royer2012a,Grienberger2017a,Zutshi2022a}. The role of EC in particular as been tested with silencing experiments and shown to be essential for instructive plasticity of CA1 \cite{Grienberger2022a}. Because connections from EC to CA1 are facilitating \cite{Jackman2016a}, EC bursting is more likely to act as an instructive signal than isolated spikes. 

As these properties of instructive forms of plasticity echo similar burst-dependent plasticity in the cerebellum \cite{Shadmehr2010a,Yang2014a,Herzfeld2015a} and other structures \cite{Mejias2013a,Muller2019a}, such that the relationship between learning and bursting in cortex is suspected and much sought-after. High-frequency bursting of long duration start to occur in cortex upon entering an associative learning task \cite{Wang2020a}. Blocking either feedforward or feedback connections blocks learning \cite{Doron2020a}. Consistently, dendrite targeting inhibition blocks learning \cite{Chen2015a,Doron2020a}. Furthermore, if bursting controls plasticity by providing a unit-specific representation of error \cite{Payeur2021a}, one would predict that the relevant period for this instructive signal to act would come after any cues are given and after reward delivery or lack thereof. An increase in burstiness has been shown to encode errors with a delay of about 1 s with respect to the cue \cite{Naud2022a}, suggesting that second-long eligibility traces similar to hippocampus may be at work in cortex. Consistently silencing pyramidal cell activity during the reward period but after the sensory cue has been shown to alter learning of a sensory association task\cite{Ford2022a}.


 \section{Discussion}

 While we have reviewed a large number of supporting evidence that the brain uses a ternary code for representing, transmitting and processing information, and that this ternary code is at play in principal cells of the thalamus, cortex, cerebellum and hippocampus,  there remains the question: is this language spoken only by these select neurons or will future scrutiny extend its boundaries? In cortex, the presence of burst-generating calcium spikes depend on neuron size, cortical layer and species \cite{Waters2003a,BeaulieuLaroche2018a,Ledergerber2010a,Fletcher2019a}. Yet even for layer 2-3 neurons that have been shown to have only weak calcium spikes \cite{Waters2003a}, these neurons engage in bursting in vivo \cite{Senzai2019a} and have burst-dependent transmission and synaptic plasticity \cite{Lefort2009a,Froemke2006a}. It is therefore more likely that other bursting mechanisms are at play in these cells \cite{Gidon2020a,Brumberg2000a,Haj1997a}. 

GABAergic interneurons of the cortex, on the other hand, may show burst-dependent transmission \cite{Campagnola2022a}, but have not been shown to have burst-dependent plasticity \cite{Hennequin2017a} or special burst-generation mechanisms. Recent work has revealed that these cells are producing NMDAR-based dendritic spikes, which can alter the ISI dispersion as in burst coding cells \cite{Friedenberger2022a}. Putative GABAergic cells have also been shown to modulate bursting in an attention task \cite{Voloh2018a} in vivo. Yet this forms an isolated body of evidence, so that the widespread nature of burst coding remains to be proven.

\section{Conclusion}

A ternary neural code is manifest from observations in some of the most studied neuron types in neuroscience. While neurons emit action potential in relative isolation by accumulating changes to the membrane potential, bursts typically arise from calcium flowing through VGCCs. We argued that the ternary code can be seen as a way for cells to communicate both elevated membrane potential and elevated levels of intracellular calcium. Synaptic plasticity being engaged by elevated calcium rather than elevated membrane potential, the ternary neural code may have arisen as a need to distinguish between these distinct intracellular signals when communicating with other cells. By signalling to other cells a signal for undergoing plasticity, neurons could engage in elaborate coordination of plasticity. Alternatively, a ternary neural code may have enabled other functions, such as the ability to manipulate and target top-down attention. The extent of burst coding in the central nervous system as well as its implications for theories of cognitive processes still need much to be discovered. For this, in vivo recordings must retain a sensitivity to both single spikes and burst, which remains a challenge for current high-throughput methodologies.

\end{document}